# Fabrication and Imaging of Monolayer Phosphorene with Preferred Edge Configurations via Graphene-Assisted Layer-by-Layer Thinning

*Yangjin Lee[1,2], Sol Lee[1,2], Jun-Yeong Yoon[1,2], Jinwoo Cheon[2,3,4], Hu Young Jeong[5,*], and Kwanpyo Kim[1,2,*]*

[1]Department of Physics, Yonsei University, Seoul 03722, Korea.

[2]Center for Nanomedicine, Institute for Basic Science (IBS), Seoul 03722, Korea.

[3]Graduate Program of Nano Biomedical Engineering, Yonsei-IBS Institute, Yonsei University, Seoul 03722, Korea

[4]Department of Chemistry, Yonsei University, Seoul 03722, Korea.

[5] UNIST Central Research Facilities (UCRF) and School of Materials Science and Engineering, Ulsan National Institute of Science and Technology (UNIST), Ulsan 44919, Korea.

*Address correspondence to K.K. (kpkim@yonsei.ac.kr) and H.Y.J. (hulex@unist.ac.kr)

**Abstract:** Phosphorene, a monolayer of black phosphorus (BP), is an elemental two-dimensional material with interesting physical properties, such as high charge carrier mobility and exotic anisotropic in-plane properties. To fundamentally understand these various physical properties, it is critically important to conduct an atomic-scale structural investigation of phosphorene, particularly regarding various defects and preferred edge configurations. However, it has been challenging to investigate mono- and few-layer phosphorene because of technical difficulties arising in the preparation of a high-quality sample and damages induced during the characterization process. Here, we successfully fabricate high-quality monolayer phosphorene using a controlled thinning process with transmission electron microscopy, and subsequently perform atomic-resolution imaging. Graphene protection suppresses the e-beam-induced damage to multi-layer BP and one-side graphene protection facilitates the layer-by-layer thinning of the samples, rendering high-quality monolayer and bilayer regions. We also observe the formation of atomic-scale crystalline edges predominantly aligned along the zigzag and (101) terminations, which is originated from edge kinetics under e-beam-induced sputtering process. Our study demonstrates a new method to image and precisely manipulate the thickness and edge configurations of air-sensitive two-dimensional materials.

Black phosphorus (BP) is an elemental two-dimensional material with unique properties including strong layer-dependent electronic bandgap modulation,[1, 2] high charge carrier mobility,[3, 4] and exotic in-plane anisotropic physical properties.[5-12] In particular, significant research has been conducted on the monolayer form of BP, phosphorene, using both theoretical and experimental methods.[8, 9, 13, 14] However, experimental studies on ultra-thin BP have been limited because of sample degradation from ambient exposure/characterization process and the preparation of pristine monolayer phosphorene being extremely challenging.[14-18] Although some studies on the optical and electrical properties of monolayer phosphorene have been reported, atomic-scale imaging of pristine phosphorene and defects therein[19, 20] have not been studied in detail so far. This has hindered the fundamental understanding of the physical properties and structural stability of phosphorene.

The atomic scale characterization and structural modification of BP using various microscopic techniques, including transmission electron microscopy (TEM) and scanning tunneling microscopy (STM), have been reported.[21-28] However, these studies have mainly investigated somewhat thick BP samples (samples that are a few-layer or thicker), and analysis of the atomic-scale structure and defects of monolayer and bilayer phosphorene are yet to be reported. Moreover, theoretical calculations suggest that phosphorene has a relatively low knock-on damage threshold for TEM imaging.[29] Therefore, to image monolayer and bilayer phosphorene samples via TEM, two challenges need to be overcome: preparation of ideal ultra-thin phosphorene samples and mitigation of e-beam-induced damages.

Herein, we report an atomic-scale analysis of ultra-thin phosphorene performed using aberration-corrected TEM. We successfully fabricated monolayer and bilayer phosphorene via a controlled thinning process under TEM, and subsequently performed atomic-resolution imaging. Graphene lamination from one-side induces the controlled thinning of BP samples, rendering high-quality monolayer and bilayer regions. The layer-number-dependent TEM experiments were compared with TEM image simulations, confirming the identification of sample thicknesses in ultra-thin regions. We also observed the formation of crystalline edge predominantly along the zigzag and (101) terminations from e-beam-induced etching. These preferred edge configurations can be explained by edge kinetics under e-beam-induced atomic sputtering process. The present study provides a novel method to precisely control the number of layers and edge terminations of easily degradable 2D materials at atomic resolution.

BP/graphene vertical heterostructures were prepared inside a globe-box via the dry transfer method, which allows for the fabrication of pristine samples without exposure to air (see Methods and Supporting Figure S1 for detailed information). Our previous study indicates that the optical transmittance of BP samples reduces by 3.3 % per layer.[25] For example, as shown in Figure 1b, a thin part of a BP sample on a PDMS substrate (Figure 1a) shows a transmittance reduction of approximately 10 %, which is the expected value for a trilayer BP sample. Subsequently, the identified BP samples were stamped and directly transferred to holey $Si_3N_4$ TEM grids. The identical identification and transfer methods were also used to transfer single and few-layer graphene to the TEM grids. By aligning the flakes under an optical microscope during the transfer process, the vertically stacked BP/graphene TEM samples were prepared as shown in Figure 1c. Figure 1d shows a high-resolution TEM image of a fabricated 3-layer BP/graphene sample, which exhibits a high-quality BP lattice structure with minimum surface residues and clean interface. The Fourier transform of the image also confirms the sample vertically stacked with graphene and BP.

TEM samples of three different configurations were compared in our study as shown in Figure 2a and 2b: 1) trilayer BP samples, 2) trilayer BP samples protected on one-side with graphene from top or bottom (namely G/BP or BP/G), and 3) trilayer BP samples encapsulated on the top and bottom by graphene (G/BP/G). TEM images of these prepared samples displayed high-quality BP crystal structures (Supporting Figure S2). To investigate the e-beam irradiation effect on the three different TEM samples, high-resolution imaging was performed under identical imaging conditions in terms of vacuum level, dose rate, and magnification. The samples were monitored until the BP crystals were significantly damaged. To quantify the e-beam-induced damages to the samples, the BP diffraction signals from fast Fourier transforms (FFT) of time-series TEM images were inspected as a function of the total accumulated dose ($e/nm^2$). At the initial imaging stage, clear diffraction signals from both BP and graphene were found and can be labeled as shown in Figure 2b.

This study indicates that the graphene protection strongly suppresses the e-beam-induced structural damages to BP. Figure 2c displays the intensity of BP (101) diffraction spot as a function of total accumulated electron dose. The (101) diffraction signal from the bare BP sample disappeared because of e-beam-induced structural degradation at an accumulated dose of ~$1.7\times10^7$ $e/nm^2$. One-side protection by graphene from bottom (BP/G sample) suppress the e-beam-induced damage to BP by approximately 3 times, displaying the diffraction peak up to an accumulated dose of ~$5.5\times10^7$ $e/nm^2$. The similar level of dose tolerance for BP samples can be also obtained through protection by top graphene (G/BP configuration). Moreover, encapsulating the BP using graphene on both sides further suppresses the beam damage, and the (101) peak was sustained up to an accumulated dose of ~$2.5\times10^8$ $e/nm^2$. We confirmed that other BP diffraction peaks displayed similar accumulated-dose-dependent behaviors (Supporting Figure S3). Figure 2d shows the evolution of (101) BP signals during the structural degradation at a couple of selected accumulated electron doses.

The suppressed e-beam-induced damage through graphene encapsulation is consistent with previously reports on other 2D materials protected by graphene.[30-32] The e-beam-induced damage during the TEM observation can be mainly divided into two categories: the knock-on damage and radiolysis.[33] The latter can also lead to chemical etching for 2D materials, which results from reactions with surface contaminants and residual gaseous species in a microscope.[34] Considering the thermal conductivity of BP ($\sim$10 $Wm^{-1}K^{-1}$)[35] and plasmon excitation ($<E> \sim$ 20 eV)[24] as a main inelastic scattering contribution, we can safely ignore the heating effect from e-beam irradiation[33] under the experimental conditions. When a specimen is in direct contact with graphene, graphene can mitigate the radiolysis of the specimen via a fast charge transfer. On the other hand, knock-on damage can still occur, even for graphene-protected BP samples. From our observation that the G/BP/G sample has $\sim$13 times higher durability compared with the bare BP sample, we can conclude that radiolysis is the main damaging mechanism for bare BP samples. The vacuum level during TEM operation can be an important factor regarding the radiolysis as demonstrated by works performed under ultra-high vacuum as well as controlled atmospheres.[36, 37]

BP with graphene protection on one-side displays controlled layer-by-layer etching under the e-beam irradiation (Figure 3a). A close observation of peak intensity for BP/G sample in Figure 2c indicates that there is an intermediate intensity step at the electron dose between $3\times10^7$ and $5\times10^8$ e/nm$^2$ during the intensity reduction, which implies a step-wise degradation process. This intermediate intensity results from the layer-by-layer etching of BP, which was also confirmed by high-resolution imaging as shown in Figure 3. Figure 3b-d show snapshots of the TEM images, which display the layer-by-layer etching from trilayer to bilayer BP. The trilayer region changes its phase-contrast image pattern, and eventually another phase-contrast pattern originating from the bilayer phosphorene emerged. (Figure 3b-3d, Supporting Movie S1).

The observed layer-by-layer etching process is also consistent with the previously-discussed damaging mechanisms for BP. The schematic of the layer etching process is shown in Figure 3a. The exposed BP layer without graphene protection predominantly suffers from radiolysis damage and preferential etching proceeds from this direction. By contrast, we observed amorphization, rather than the layer-by-layer etching process, in the G/BP/G samples. A similar etching process was observed for bilayer BP/G samples (Supporting Figure S4), where high-quality monolayer phosphorene regions with considerable areas were produced, as shown in Figure 3e. The observed monolayer phosphorene clearly shows a highly crystalline structure with scattered isolated vacancies. The underlying graphene substrate can be also uncovered from monolayer phosphorene regions under the prolonged e-beam irradiation as shown in Supporting Figure S5. To confirm the samples' local layer numbers, we performed TEM image simulations and compared them with observed experimental images. For bilayer and trilayer BP structures, we assumed the conventional AB stacking.[24, 25] Figure 3f shows the atomic resolution TEM images (left panel in Figure 3f) and simulated TEM images (right panel in Figure 3f) of mono-, bi-, and trilayer phosphorene. We observed a good agreement between the simulation and experimental results. Figure 3g and 3h display the contrast intensity profiles from the regions marked in Figure 3f; the experimentally measured intensity profiles also agree well with those from simulation results. Although BP etching through plasma treatment and controlled oxidation has been reported, the direct atomic-scale imaging of monolayer after the etching process has not been previously reported.[38-41]

Well-defined high-quality edge terminations of phosphorene were observed during the controlled etching process. For example, the snapshots during the etching (Figure 3b) show that edge termination predominantly separates two distinct regions along the (101) crystal termination of BP. Figure 4a also shows an atomic resolution TEM image of a well-defined step-edge between bilayer and trilayer BP regions with a zigzag edge configuration. The observed phase-contrast TEM images were compared with simulation results, which confirm that no apparent edge reconstruction occurred at the atomic scale. We also frequently observed the crystalline edge structure with (101) termination as shown in Figure 4b. Theoretical calculations suggest that the physical and chemical properties of phosphorene systems strongly depend on their edge configurations[42, 43] and we believe that the demonstrated manipulation of edge terminations via e-beams could be potentially useful to control the properties of phosphorene. We note that the thinning and crystalline edge formation were observed from both BP/G (Figures 3b-d and 4a) and G/BP (Figures 3e and 4b-c) configurations.

To study the observed preferred edge terminations of phosphorene in detail, we analyzed the time series of TEM images under e-beam stimulation (Supporting Movie S2). Figure 4c shows time series TEM images of mainly bilayer phosphorene with a partially exposed monolayer central pit. Under e-beam stimulation, the monolayer region gradually increases its size over time and the exposed edge frequently changes its local configurations. The terminated edges were predominantly aligned along the zigzag or (101) terminations, but the armchair edge configuration was rarely observed (Figure 4c and 4d). We recorded the edge population over time and studied the dynamical evolution (Figure 4f). The average fraction of the zigzag, (101), and armchair orientations were 48.3 %, 42.1 %, and 9.6 %, respectively.

The preferred edge formation can be explained by kinetics for removing a phosphorus atom from different edge configurations. Although the radiolysis is still in effect, the knock-on damaging mechanism will also play an important role during the thinning process of monolayer and bilayer phosphorene, because the radiolysis is largely suppressed by graphene protection. Das et al.[23] has calculated the energy barriers to remove a phosphorus atom from different edge configurations, which shows that the barrier is the lowest for the armchair edge configuration. Therefore, the armchair edge can easily transform to another edge configuration via the atomic sputtering process during TEM imaging. This different edge energetics can be also inferred from the puckered structure of phosphorene. The atoms in the puckered phosphorene crystal can be divided into atoms in the top layer (green color) and bottom layer (purple color) as shown in Figure 5a. In-plane bond length (2.22 Å) is slightly shorter than the out-of-plane bond length (2.24 Å). Therefore, the removal of a phosphorus atom by breaking an in-plane bond will be more difficult.[22, 44] Considering the different edge bond configurations, we expect that the edge atoms in the zigzag configuration will be more stable because the atoms at the zigzag termination possess two in-plane bonds. As atoms in the armchair configuration possess one in-plane and one out-of-plane bond, the energy cost for atomic sputtering will be lower compared with that of zigzag edges.

Monte Carlo simulations were performed based on this different edge energetics to simulate the atomic sputtering process as shown in Figure 5. From a single vacancy state, we removed phosphorus atoms, one by one, from the vacancy edge and simulated the process of vacancy growth. We categorized edge atoms into three types: type-1 atoms unstable possessing only single neighbor, typy-2 atoms at armchair termination, and type-3 atoms at zigzag termination (Figure 5a). Type-1 atoms with one neighbor have priority for sputtering process and they were chosen as a sputtering candidate for the next atom removal process. Simulations under five different probability ratios for armchair (type-2) and zigzag (type-3) atom removals were considered. It was evident that our experimental observation was reproduced in cases of higher probability to remove armchair atoms compared with zigzag, which is consistent with our expectation of higher stability of zigzag edge termination (Figures 5b-f). Among different sputtering probability, the experimentally observed edge fraction and the aspect ratio of hole shape (Supporting Figure S6) were reproduced from simulation results from 2:1 sputtering probability as shown in Figure 5b and Supporting Figure S7. The elastic scattering theory also predicts the sputtering probability ratio approximately 2:1 as shown in Supporting Figure S8.[45, 46] It is noteworthy that our result is the first clear observation of a well-defined crystalline edge termination of phosphorene at the atomic resolution. The higher stability of the zigzag edge configuration may be related to a recently reported phosphorene nanoribbon formation with the zigzag edge orientation.[47]

In conclusion, we applied graphene as a protection barrier to mitigate the radiolysis damage and demonstrated the remarkably increased stability of phosphorene under e-beam irradiation. We also demonstrated the fabrication of a high-quality monolayer phosphorene under TEM using a graphene-assisted layer-by-layer etching. The phosphorene/graphene sample geometry allowed us to observe the ideal edge configurations, which are consistent with structures from theoretically calculated edge kinetics under e-beam sputtering. We envisage this technique could be employed for detailed studies of defects, edge structures, and polymorphs[48] in monolayer phosphorene and other beam-sensitive materials.

**Methods**

**TEM Sample Preparation.** TEM samples were prepared using the polydimethylsiloxane

(PDMS)-based dry transfer method.[49] PDMS films were fabricated using a 10:1 (w/w) pre-polymer and curing agent mixture (Sylgard 184, Dow Corning Co.). BP (purchased from Smart Elements) flakes were prepared on the PDMS using mechanical exfoliation. Three-layer BP flakes were identified using an optical microscope under transmission mode.[25] Optical microscope images were acquired using a Leica DM-750M with visible light. To minimize BP surface degradation, all sample fabrication and optical characterization processes were performed inside a nitrogen-filled glove box (oxygen concentration was maintained below 0.2 ppm). Identified trilayer BP flakes were transferred to holey $Si_3N_4$ TEM grids (purchased form Nocada Inc.) using the optical microscope equipped with micro-manipulators. In the case of heterostructures (BP/G and G/BP/G), mechanical exfoliation and transfer of graphene flakes were performed using the same process as above. To minimize the charging effect during TEM imaging, the $Si_3N_4$ TEM grids were coated with 3 nm of carbon using PECS (Gatan, USA) prior to 2D flake transfer.

**TEM imaging and image simulation.** TEM images and FFT patterns were acquired using an FEI Titan[3] G2 operated at 80 kV, which is equipped with double Cs-aberration corrector and monochromator. The TEM vacuum level near the samples was $7\times10^{-8}$ torr. Preliminary diffraction data were acquired using a FEI Tecnai F-20 at 200 kV. For TEM time series, an exposure time of 0.2 seconds along with a processing time of 0.5 seconds was used. TEM image simulations were performed using MacTempas software which implements the Multislice calculations for HRTEM imaging. The atomic potential was calculated by a sum over independent atomic potentials. TEM imaging acquisition conditions including an acceleration voltage value of 80 kV, a Cs value of -13 µm, a convergence angle of 0.1 mrad, and a defocus value of 7 nm were used for simulations. A mechanical vibration of 0.5 Å without adding extra aberration was used. Atomic structures used for the TEM simulation were modeled using the Virtual NanoLab program.

**Calculations.** Monte Carlo simulations were performed using in-house MATLAB codes. From a single vacancy state, 100 atoms were removed, atom by atom, from candidates located at exposed phosphorene edge. Upon creation, a type-1 atom was chosen as a next candidate and removed immediately from an exposed edge configuration. Five different probability ratios for armchair (type-2) and zigzag (type-3) atom removals were considered. The considered ratios were 4:1, 2:1, 1:1, 1:2, and 1:4. For each ratio, the simulations were performed 10 times and the average fraction of edge terminations and the aspect ratio of hole shape (zigzag/armchair directions) were investigated. The calculations of probability for edge atom sputtering were performed using the elastic scattering theory. The calculation details can be found in the Supporting Information.

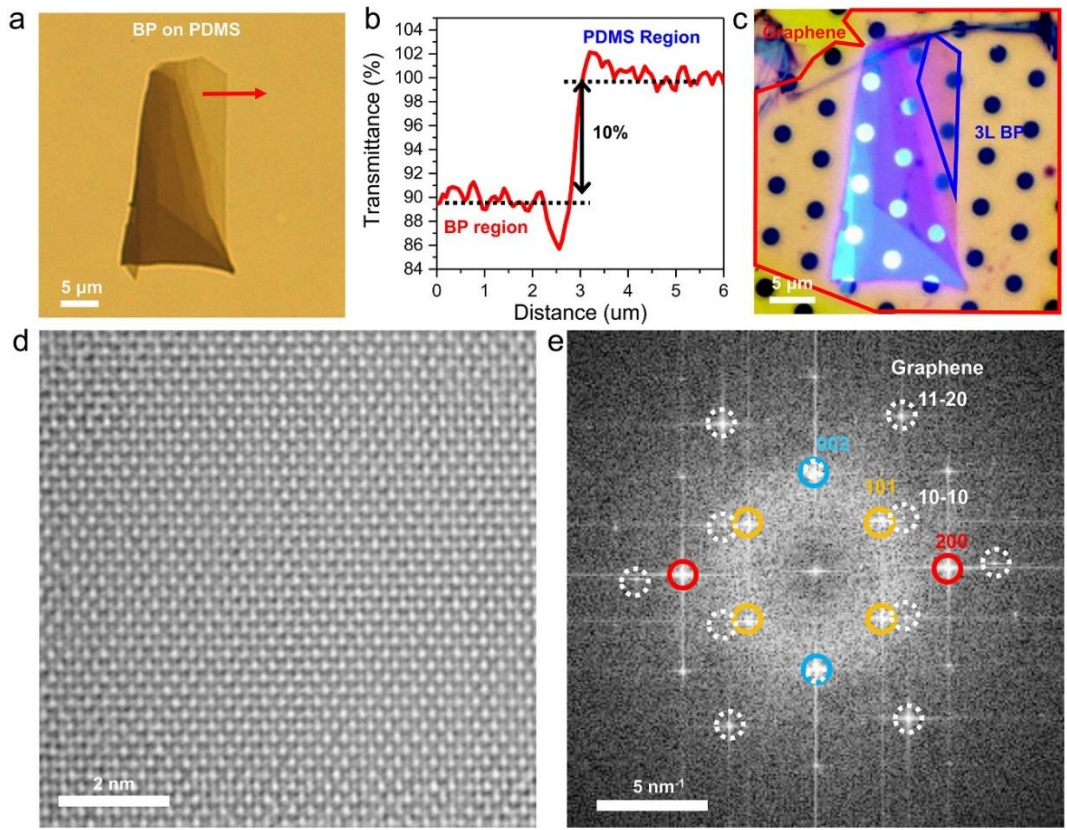

**Figure 1.** Black phosphorus (BP)/graphene vertical heterostructure. (a) Optical image of a black phosphorus flake on PDMS under transmittance mode. (b) Measured optical transmittance along the red line in panel (a). (c) Optical image of a BP/graphene TEM sample. Trilayer BP region is highlighted. (d) HR-TEM image of trilayer BP/graphene. (e) FFT signal of observed image in panel (d). Graphene diffraction signals are marked with white dashed circles. (002), (101), and (200) BP diffraction peaks are marked with blue, orange, and red colored dots, respectively.

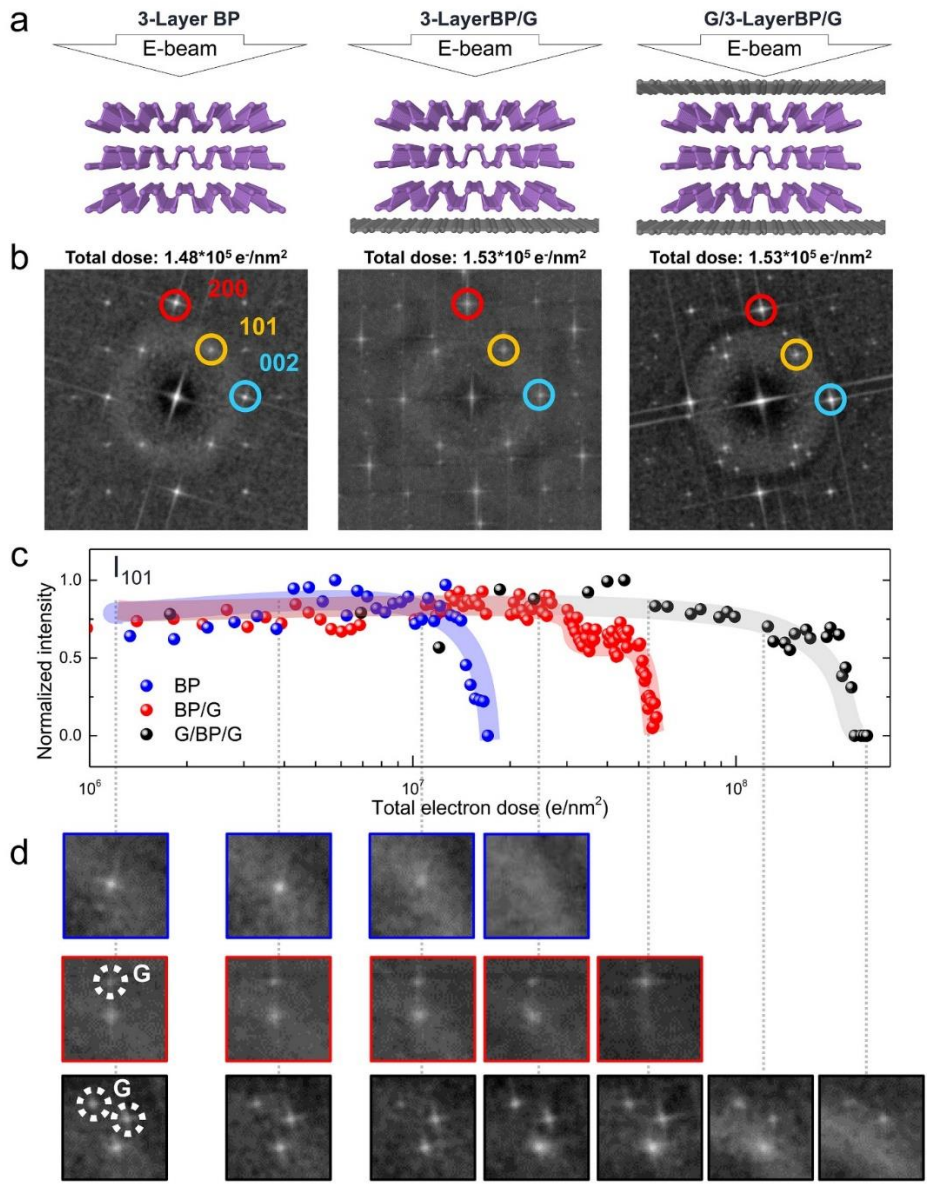

**Figure 2.** Reduced e-beam-induced damage of phosphorene by graphene protection. (a) Three different sample configurations for control experiments. (b) FFT patterns of three configurations at the initial imaging stage. (c) Evolution of BP (101) peak intensity as a function of total accumulated electron dose. (d) Zoom-in FFT images of (101) peak at selected total electron dose. Nearby graphene signal is marked by dashed white circles. (Blue box: BP, Red box: BP/G, Black box: G/BP/G)

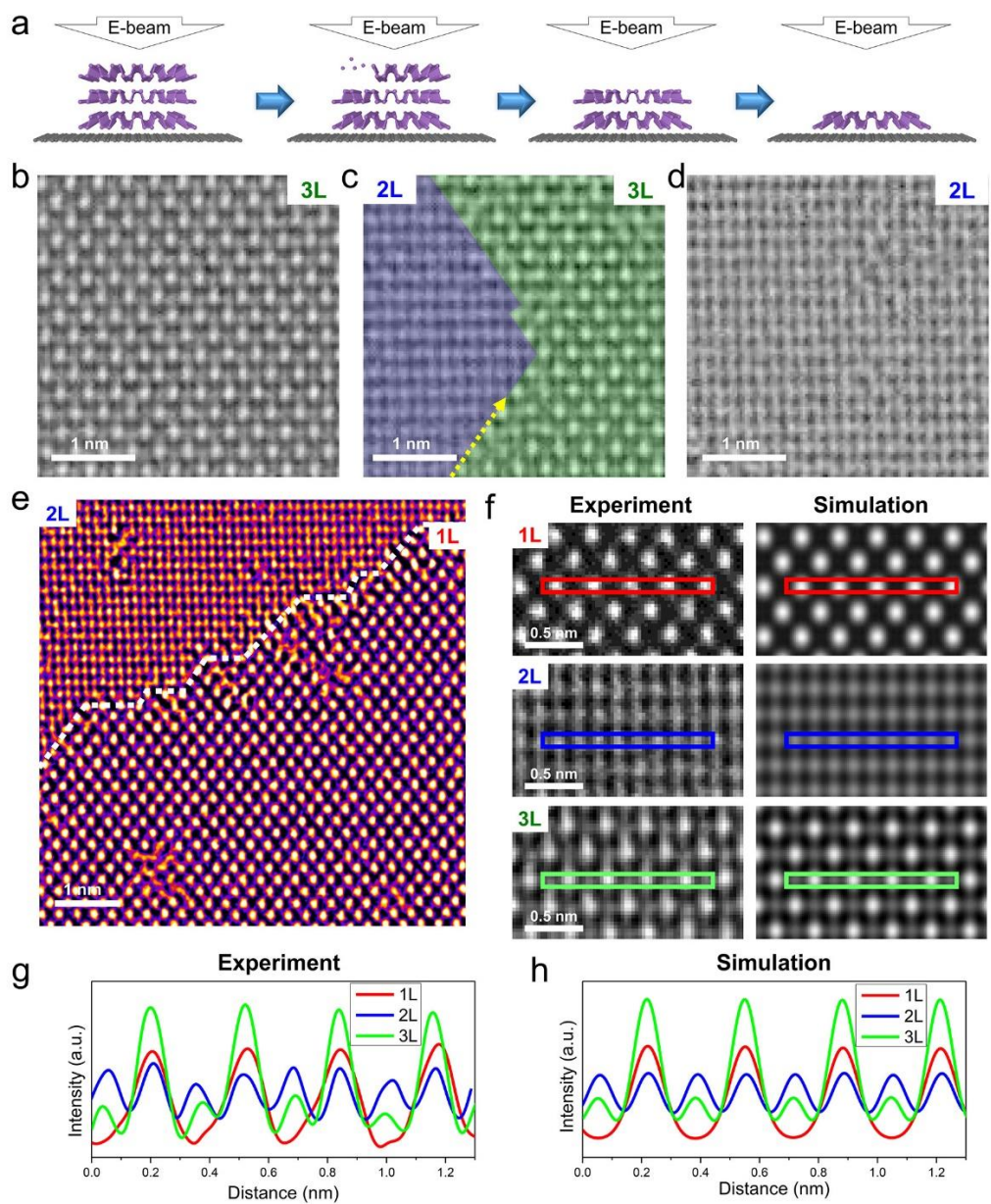

**Figure 3.** Controlled etching of phosphorene to monolayer. (a) Schematic illustration of e-beam-induced etching process. (b-d) Time series of HR-TEM images showing the thinning from trilayer to bilayer. (e) HR-TEM image showing bilayer and monolayer phosphorene. Color-code fire of ImageJ program was used. (f) Experimentally observed HR-TEM image and simulated image for monolayer, bilayer, trilayer. (g) Experimentally measured intensity profile from panel f. (h) Simulated intensity profile.

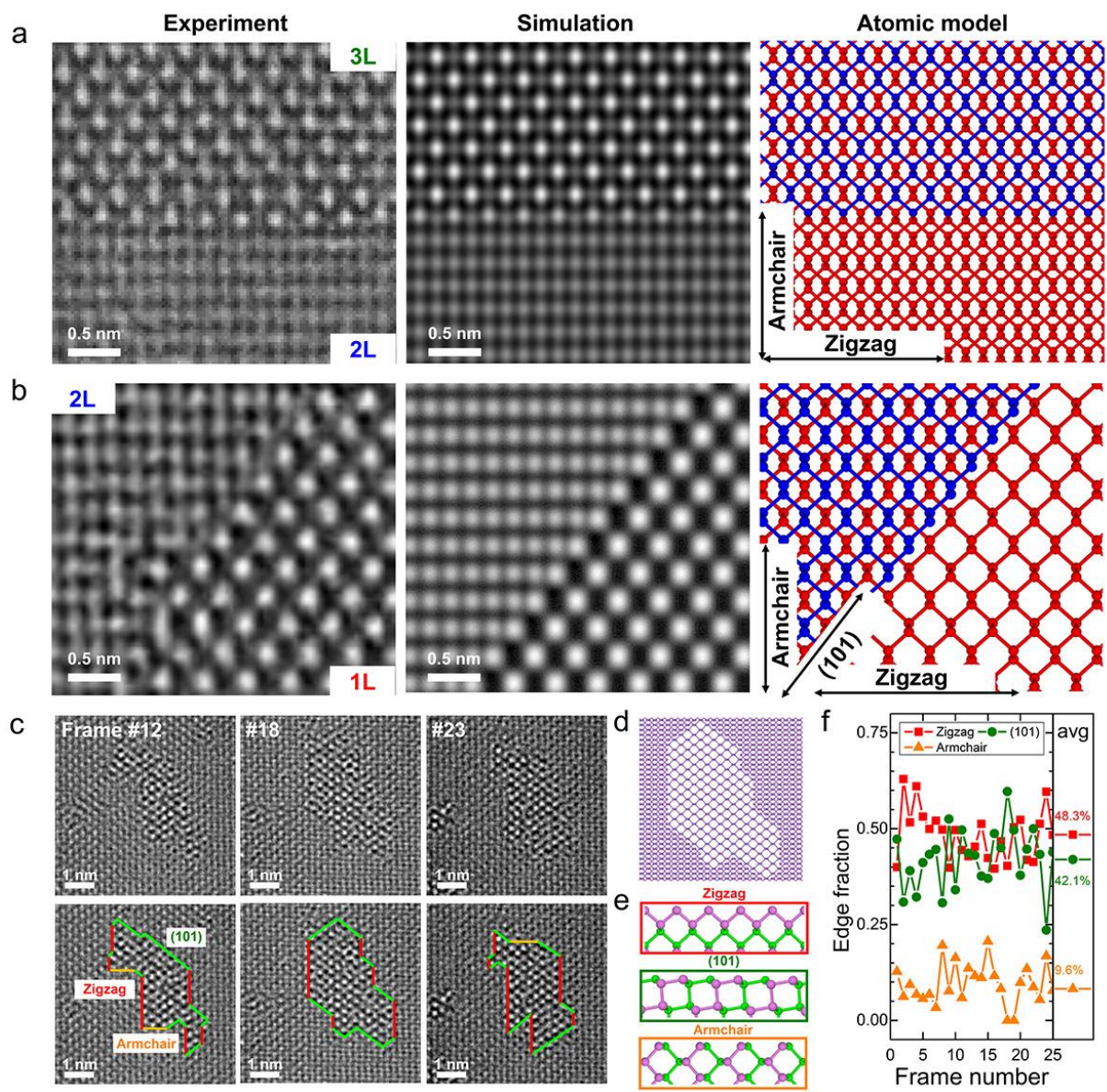

**Figure 4.** Preferred crystalline phosphorene edge terminations. Experimental HR-TEM image, simulation image, and structure model of step edges of phosphorene along (a) zigzag direction and (b) (101) termination. (c) Selected snapshots from TEM time series of a monolayer domain in bilayer surrounding. (d) Atomic model of a monolayer pit in bilayer phosphorene. (e) Atomic models of zigzag, (101), and armchair edge structures. (f) Time-evolution of observed edge fraction from the time series.

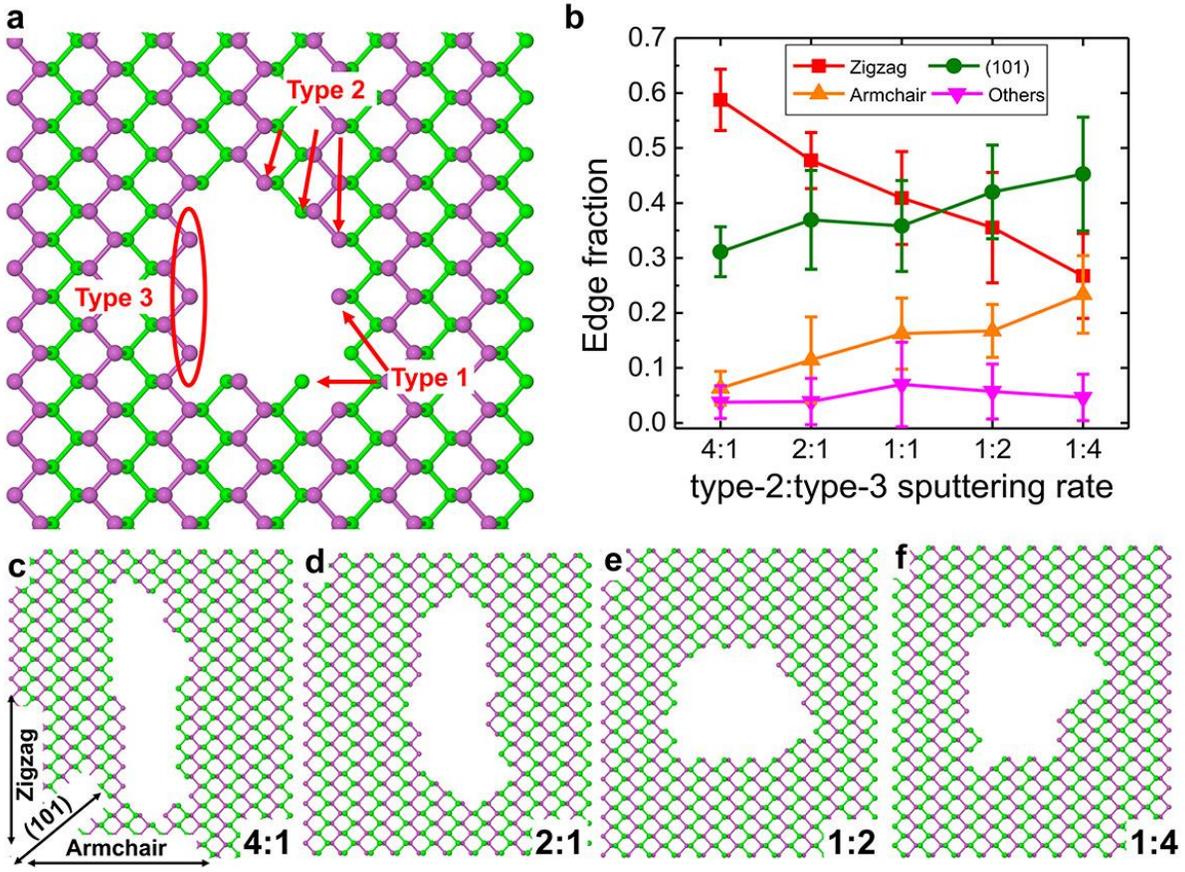

**Figure 5.** Monte Carlo simulations of atom sputtering process from edge. (a) Categorization of edge atoms depending on bond types. Type-1 atoms possess single neighbor atom. Type-2 edge atoms have one neighbor in the same atomic plane and one in the difference atomic plane. Type-3 edge atoms have two neighbors in the sample atomic plane. (b) Edge termination fraction as a function of sputtering probability ratio between type-2 and type-3 atoms. The error bars represent the standard deviations from 10 simulations for each probability ratio. (c-f) Monte Carlo simulation results after removal of 100 atoms under sputtering probability ratios of 4:1, 2:1, 1:2, and 1:4, respectively.

## ASSOCIATED CONTENT

**Supporting Information**. The Supporting Information is available free of charge on the ACS Publications website.

Calculations of differential cross-section for edge atom sputtering by elastic scattering theory, schematic of TEM sample fabrication process, extra optical images, extra TEM characterization data of samples, extra Monte Carlo simulation results, and TEM movies showing the thinning process and preferred crystalline phosphorene edge formation.

## Corresponding Author

E-mail: kpkim@yonsei.ac.kr and hulex@unist.ac.kr

## Author Contributions

The manuscript was written through contributions of all authors. All authors have given approval to the final version of the manuscript.

## Notes

The authors declare no competing interests.

## Acknowledgements

This work was mainly supported by the Basic Science Research Program through the National Research Foundation of Korea (NRF-2017R1A5A1014862, NRF-2018R1A2B6008104, and NRF-2019R1C1C1003643), the Institute for Basic Science (IBS-R026-D1), and Future-leading Research Initiative 2018-22-0054 of Yonsei University. K.K. also acknowledges support from the POSCO Science Fellowship of POSCO TJ Park Foundation.

**Table of Contents**

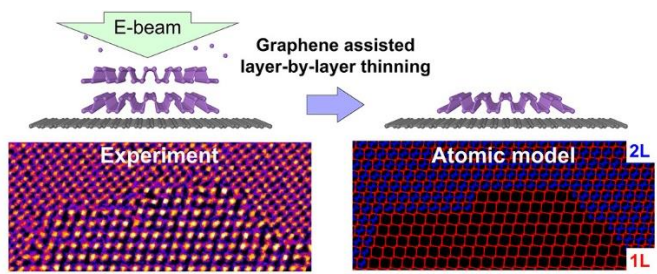